\documentclass[
    ,final            
  ]
  {aipproc}

\layoutstyle{8x11double}
\usepackage{latexsym}
\usepackage{bm}
\usepackage{amsmath,amssymb}
\usepackage{mathrsfs}
\usepackage{tabularx}
\usepackage{float}
\usepackage{enumerate}

\begin{document}

\title{Axionic Mirage Mediation}


\author{Shuntaro Nakamura}{
  address={Department of Physics, Tohoku University, Sendai, 980-8578, Japan}
}

\author{Ken-ichi Okumura}{}

\author{Masahiro Yamaguchi}{}

\begin{abstract}
In this talk, we propose a model of mirage mediation,  in which
Peccei-Quinn symmetry is incorporated. In this \textit{axionic mirage
mediation}, it is shown that the Peccei-Quinn symmetry breaking scale is dynamically determined 
around $10^{10}$ GeV to $10^{12}$ GeV due to the supersymmetry breaking effects. 
The problems in the original mirage mediation such as the $\mu$-problem and the moduli problem can be solved simultaneouly. 
Furthermore, in our model the axino, which is the superpartner of the axion, is the lightest sparticle. 
\end{abstract}

\maketitle


\section{Introduction}
 Supersymmetry (SUSY) deeply fascinates us as a solution to the hierarchy problem. 
It has, however, to be broken because superparticles have not been discovered yet. 
The phenomenological aspects such as mass spectrum, collider signature and so forth, depend on how SUSY is broken. 
Therefore, to understand the SUSY breaking mechanism is important. 
In superstring theory, one of the most plausible mediation mechanisms is the moduli mediation. 
On the other hand, the mediation associated with the super-Weyl anomaly, the anomaly mediation, is also inevitable. 
Recently, KKLT \cite{Kachru:2003aw} have proposed the interesting set-up to stabilize the modulus with the relatively heavy mass by considering some non-perturbative effects, and in this and similar set-ups, the anomaly mediation contribution is comparable with the modulus mediation one. 
This type of mediation mechanism is often called the mirage mediation  \cite{Choi:2004sx}.
The mirage mediation is a natural mediation mechanism in the string modulus stabilization and very interesting because it can solve the tachyonic slepton problem in the pure anomaly mediation, and it has characteristic mass spectra as well.
However, it suffers from two crucial problems. One is the $\mu$-/$B \mu$-problem and the other results from cosmology. 
The former stems from the fact that $B$ parameter would easily become of the order of the gravitino mass, $m_{3/2}$, which 
is $\mathcal{O}(10)$ TeV.  
The latter leads to the result that the sizable production rate of gravitinos from the modulus decay aggravates 
the conventional moduli problem \cite{Endo:2006zj}. (See, for the case of the inflaton and the Polonyi field \cite{Asaka:2006bv}).  
Such aggravation comes from the overclosure of the neutralinos produced by the gravitinos if it is the lightest superparticle (LSP). 
In this talk, we show that axionic extension of mirage mediation, \textit{axionic mirage mediation} \cite{Nakamura:2008ey}, can solve not only the $\mu$-/$B \mu$-problem but the cosmological moduli problem simultaneously.

\section{The model}

\begin{figure}
   \includegraphics[width=5.5cm, clip]{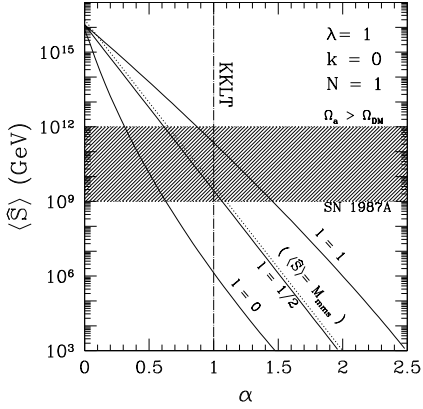} 
   \caption{The PQ scale $f_{\rm PQ}\simeq \langle \hat{S} \rangle $ in the axionic  mirage mediation. 
             Here, $\ell$ is the modular weight of the messenger. 
            } 
\label{fig:PQ scale}
\end{figure}

Let us consider the hadronic axion model in the KKLT set-up 
by introducing the axion superfield $S$ and
 $N$ pairs of messenger fields $\Psi$ and $\bar{\Psi}$.
Here, $S$ is a singlet under any unbroken gauge symmetry, $\Psi$
and $\bar{\Psi}$ are vectorlike representation of the SU(5) gauge group and 
we use $X$ to denote the modulus.
We assigned the PQ charge as $Q_{\rm{PQ}}(S) = -2$, $Q_{\rm{PQ}}(\Psi) = Q_{\rm{PQ}}(\overline{\Psi}) = 1$ and $Q_{\rm{PQ}}(X) = 0$. 
The superpotential has Yukawa coupling, 
\begin{eqnarray}
  W = \lambda \hat{S} \hat{\Psi} \hat{\bar{\Psi}} + \cdot \cdot \cdot,
\end{eqnarray}
which is allowed by the PQ symmetry. 
The hat means the absorption of the chiral compensator $\Phi$, that is $\hat{S} \equiv \Phi S$. 
$S$ is a flat direction in SUSY limit, however, it is lifted by
the SUSY breaking,  $F_{X,\Phi}$, caused by the mirage mediation. 
Here, $F_X$ and $F_\Phi$ stand for the auxiliary field of the modulus and 
that of the compensator, respectively. 
We assume that $S$ is stabilized far away from the origin, breaking the
PQ symmetry. Since this gives a huge mass for the messengers, 
we can integrate out the messengers. This leads to 
\begin{eqnarray}  \label{eq: wave function renormalization of S}
   \mathcal{L} = 
   \int d^4 \theta (X + X^\dagger)^k 
                   Z_S \left( 
                          \sqrt{ \frac{\hat{S}^\dagger \hat{S}}{\Phi^\dagger \Phi}}, \, X + X^\dagger 
                       \right) 
                   | \hat{S} |^2,
\end{eqnarray}
where $Z_S$ is the wave function renormalization of $S$ at $\mu = |\hat{S}|$ and  $k$ the modular weight 
of $S$.
According to the renormalization group equation (RGE) of $m^2_S$, 
we can find that the scalar potential of $S$ has a minimum at 
some scale $\langle \hat{S} \rangle$ \cite{Nakamura:2008ey}
. 
We depicted $\langle \hat{S} \rangle$ as a function of 
$\alpha \equiv \frac{m_{3/2}}{(F_X/2X_R) \ln(M_{Pl}/m_{3/2})}$ in fig.\ref{fig:PQ scale}.  
It is noted that the PQ scale, $f_{\rm PQ} \simeq \langle \hat{S} \rangle$, can be within, 
so-called, the axion window $10^9 {\rm GeV} \lesssim f_{\rm PQ} \lesssim 10^{12-13}$ GeV. 
In what follows, we consider the case $\langle \hat{S} \rangle \simeq 10^{10}$ GeV.


From eq.\eqref{eq: wave function renormalization of S}, we can find that the mass of the axino $\widetilde{a}$, which is the superpartner of the axion, is calculated  at two-loop order in contrast with the masses of the other fields being at one-loop. Thus it will be the LSP and candidate for the dark matter (DM).

\section{The $\mu$-/$B \mu$-problem }  \label{mu problem}
An important property of this model is that it provides a natural solution to the $\mu$-/$B \mu$-problem.
Let us consider the following superpotential and $\Omega$ function (See, for instance, \cite{Pomarol:1999ie})
\begin{gather} 
   W = y_1 T H_1 H_2 + y_2 S_1 S_2 T, \label{mu-term superpotential} \\
   \Omega = |S_1|^2 + |S_2|^2 + |T|^2 + \kappa S_1^\dagger S_2 + \rm{h.c.} , \label{mu-term f function}
\end{gather}
where we introduced new singlets $S_1$, $S_2$ and $T$, whose PQ charge is assigned as $Q_{\rm{PQ}}(S_1) = Q_{\rm{PQ}}(S_2) = -2$ and $Q_{\rm{PQ}}(T) = +4$, and $y_1$, $y_2$ and $\kappa$ are constants. 
The function $\Omega$ has a relation to the K\"ahler potential $K$ via $\Omega = -3 e^{-K/3}$. 
For simplicity, the modular weight of singlets is set to be zero.  
When $S_1$ gets a vacuum expectation value (VEV), integrating out $T$ leads to 
the $\mu$-/$B \mu$-term in canonical normaization: 
\begin{eqnarray}
   \mu \!\!\!\!&=&\!\!\!\!
       - \frac{\kappa}{(X+X^\dag)^{(q_{H_1}+q_{H_2})/2}} \, \frac{y_1}{y_2} \frac{F_{\hat{S}_1}^\dagger}{\hat{S}_1}
\label{mu} 
         \\
   B \mu \!\!\!\!&=&\!\!\!\!
             \frac{\kappa}{(X+X^\dag)^{(q_{H_1}+q_{H_2})/2}} \, \frac{y_1}{y_2}  
             \frac{F_{\hat{S}_1}^\dag}{\hat{S}_1} \nonumber \\  
            &&\times  \left[\frac{F_{\hat{S}_1}}{\hat{S}_1}+(q_{H_1}+q_{H_2})\frac{F_X}{2X_R}\right]  
\end{eqnarray}
with $q_{H_{1, 2}}$ being the modular weight of the Higgs fields $H_1$, $H_2$. 
Since they are the same order of soft masses, the $\mu$-/$B \mu$-problem can be solved. 
It is noted that in our model there is also no SUSY CP problem. 
Since the phases of two SUSY-breaking $F$-terms, ${F_\Phi}$ and ${F_X}$, are aligned \cite{Choi:1993yd},    
we find that the phase of $B$
can be rotated away simultaneously with that of the gaugino mass and $A$-term. 
Therefore, we constructed a model in which there is no $\mu$-problem as well as SUSY CP problem.

\section{Cosmology}  \label{cosmology}

In this section, we discuss the cosmological implications of our model. 
We assume that the modulus and the saxion, which is the real part of the scalar component 
of the axion superfield, are displaced from their true minima with the amplitude 
of the order of the Planck scale, and 
their coherent oscillation start before the reheating due to the inflaton decay is completed. 
In this case, we can find that it is sufficient to discuss the cosmic evolution only after 
the modulus decay because of a large amount of the entropy production. 
In what follows, we address the implications of the modulus decay in cosmology. 

\begin{figure}[t]  
   \includegraphics[width=7.5cm, clip]{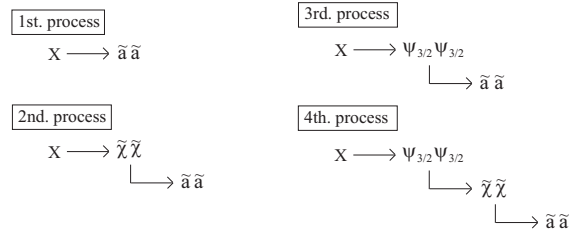} 
   \caption{The sketch of four production processes of the axino, where $\widetilde{\chi}$ denotes the NLSP. 
   } 
\label{fig: figure of decay}
\end{figure}

There are four processes producing the axino LSP via the modulus decay (See, fig.\ref{fig: figure of decay}). 
However, the dominant contribution to the axino abundance is given by the \textit{4th. process}.  
We investigated the relic abundance of the axino in various cases with the NLSP being
 bino, higgsino, stau, stop and wino. 
The result is summarized in the table.1. 
Here, case A and B represent that the annihilation rate of the NLSP at the gravitino decay is 
negligible and significant, respectively. 
It is noted that when the annihilation process of the NLSP is not effective, the relic abundance of the axino 
becomes the same value in any NLSP cases.

\begin{table}
\begin{tabular}{ccc}
\hline
   \tablehead{1}{r}{b}{NLSP}
  & \tablehead{1}{r}{b}{Case A: $\Gamma \gg n \langle \sigma v \rangle$}
  & \tablehead{1}{r}{b}{Case B: $\Gamma \ll n \langle \sigma v \rangle$} \\
\hline
$\widetilde{B}$ & $Y_{\widetilde{a}} \simeq 4.3 \times 10^{-9}$ & Does not occur  \\
$\widetilde{h}$ &  &    \\
$\widetilde{\tau}$ &   &    \\
$\widetilde{t}$ &  & Abundance is too small  \\ 
$\widetilde{W}$ &  & $Y_{\widetilde{a}} \simeq 3.3 \times 10^{-11}$   \\
\hline
\end{tabular}
\caption{Table of the abundance of the axino in the \textit{4th. process}.}
\label{tab:a}
\end{table}

\section{Results}
\begin{figure} 
   \includegraphics[width=5.0cm, clip]{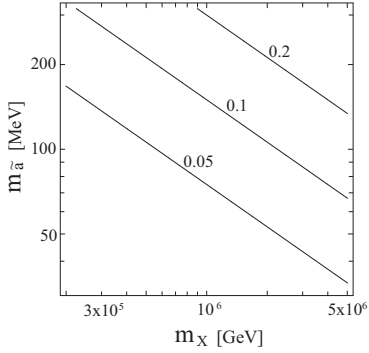} 
   \caption{Contours of the density parameter of the axino, $\Omega_{\widetilde{a}} h^2$. 
             } 
\label{fig: contour of axion with modulus mass}
\end{figure}

\begin{figure}
   \includegraphics[width=5cm, clip]{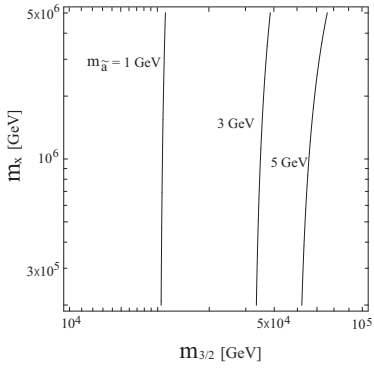} 
   \caption{Contours of the axino mass which satisfy $\Omega_{\widetilde{a}} h^2 = 0.1$. 
   } 
\label{fig: contour of axino in wino NLSP case}
\end{figure}

We are now ready to discuss whether the axino relic abundance accords with the present DM abundance. 
In fig.\ref{fig: contour of axion with modulus mass}, we plotted the axino mass density in terms of $m_X$ and $m_{\widetilde{a}}$. 
From the WMAP three year results \cite{Spergel:2006hy}, the DM abundance in the present Universe is $\Omega_{\rm{DM}} h^2 = 0.105^{+0.007}_{-0.013}$ (68 $\%$ C.L.). 
Thus,  we find from fig.\ref{fig: contour of axion with modulus mass} that the axino with $m_{\widetilde{a}} \simeq \mathcal{O}(100) $ MeV can explain the present DM abundance in any NLSP cases, if the decay of the NLSP is more effective than the annihilation in the $\textit{4th. process}$. 
In fig.\ref{fig: contour of axino in wino NLSP case}, we plotted axino mass contours which satisfy $\Omega_{\widetilde{a}} h^2 = 0.1$ in $m_{3/2}-m_X$ plane for the wino NLSP case, where we set $\lambda = N =1$, $k = 0$, and $m_{\widetilde{W}} =$ 100 GeV.  
If the wino is the NLSP and their annihilation is effective, fig.\ref{fig: contour of axino in wino NLSP case} shows that the right amount of DM can be explained by the axino with a few GeV mass. 

\section{Conclusion}

We have discussed the axionic extension of mirage mediation, \textit{axionic mirage mediation}, to remedy crucial problems in the mirage mediation: the $\mu$-problem and the moduli problem.     
It was also found that the SUSY CP problem is absent in our model.
Furthermore, we found that the axino with mass of $\mathcal{O}(100)$ MeV -- $\mathcal{O}(1)$ GeV 
can constitute the DM of the Universe.

Finally we would like to briefly mention signatures of neutralino NLSP decays at collider experiments.
In our model, the axino LSP couples to the Higgs multiplets at tree level, and hence the lifetime of the 
neutralino NLSP, if it is higgsino-like, is $\mathcal{O}(10^{-9})$ sec. when $\langle \hat{S} \rangle \simeq 10^{10}$ GeV. 
We expect that the NLSP decay into the Higgs boson will provide a spectacular signal of displaced vertex emitting hard jets, if the decay vertex can be reconstructed.


\begin{theacknowledgments}
The work was partially supported by the grants-in-aid from the Ministry of Education, Science, Sports, and Culture of Japan, No.16081202 and No.17340062. 
K.O. is supported by the Grand-in-aid for Scientific Research No.19740144
from the Ministry of Education, Culture,
Sports, Science and Technology of Japan.
\end{theacknowledgments}



\bibliographystyle{aipproc}   


\end{document}